# Inflation as the unique causal mechanism for generating density perturbations on scales well above the Hubble radius

Andrew R. Liddle

*Astronomy Centre, School of Mathematical and Physical Sciences,*
*University of Sussex, Brighton BN1 9QH.   U. K.*
arl@starlink.sussex.ac.uk

(October 27, 1994)

An examination is made of the widely held belief that inflation is the only possible causal mechanism capable of generating density perturbations on scales well in excess of the Hubble radius. A simple proof is given, which relies only on the assumption that our understanding of the universe from nucleosynthesis onwards is correct. No assumption of the underlying gravitational theory is necessary beyond that it is a metric theory.

## I. INTRODUCTION

The present popularity of the inflationary cosmology [1] is almost entirely vested in its ability to generate a spectrum of density perturbations which may lead to the formation of structure in the universe. Indeed, it is apparently widely believed amongst cosmologists that inflation is the *only* way in which density perturbations can be generated on scales significantly larger than the Hubble radius without breaching causality. Despite this, little effort has gone into determining the precise conditions under which this belief is supposed to be true. For instance, claims that the observations of microwave background anisotropies by the COBE satellite [2] support inflation have rested on the consistency of the anisotropies with the Harrison-Zel'dovich power spectrum predicted by the simpler inflationary models, rather than on the very existence, or otherwise, of perturbations on scales in excess of the Hubble radius.

The closest discussion to the present one is that by Hu, Turner and Weinberg [3], who discussed whether or not inflation may be the unique way of solving the horizon and flatness problems. The resolution of the horizon problem is intimately related to the ability to generate density perturbations, and so this paper treads similar grounds. However, the proof offered here is considerably simpler, closes loopholes in their arguments, and is more widely applicable as it generalizes simply to any metric theory of gravity.

The crucial ingredient required to prove that inflation is the only causal means of generating large scale perturbations is that the evolution of the universe follows the standard hot big bang for a reasonable portion of its recent history. The simplest assumption which supplies this is that standard nucleosynthesis, one of the cornerstones of modern cosmology [4], is correct, and that we therefore understand the universe well from that time onwards.

An examination of this result is very timely, because large scale structure observations are increasingly pointing towards models of structure formation which rely on an initial spectrum of density perturbations which extends out to scales well beyond our present observable universe [5]. One of the most direct ways in which large scale perturbations may be seen is in anisotropies in the cosmic microwave background radiation; the COBE satellite [2] has detected anisotropies on all angular scales up to the quadrupole. In the standard picture, last scattering occurred at a redshift of around 1000, with the Hubble radius at that time subtending an angle of only around one degree (somewhat less if the universe is open rather than flat). In the most popular models for large scale structure formation, the anisotropies directly sample irregularities in the distribution of matter at the time of last scattering. The COBE observations are consistent with a scale-invariant spectrum of density perturbations extending up to scales well in excess of our present observable universe, and yet more dramatically in excess of the Hubble radius at the time the microwave radiation was released.

A standard motivation for supposing that there exists a spectrum of perturbations extending to such large scales lies in the theory of cosmological inflation [1], which posits a period of accelerated expansion during the earliest stages of the universe's existence. Inflation was introduced [6] in order to solve problems connected with the initial conditions for the hot big bang model, the horizon and flatness problems. However, its greatest strength is that it provides a mechanism — the dramatic stretching and 'freezing in' of quantum fluctuations [7] — capable of generating large scale density perturbations. In the simplest models these are adiabatic (that is, genuine fluctuations in the total energy from point to point), though it is also possible that they can be isocurvature [8] (fluctuations in the relative amounts of different types of matter, *eg* nonrelativistic matter and radiation, leaving the total energy density constant).

The evidence that there were indeed perturbations on scales larger than the Hubble radius at the time the microwave background originated is not yet completely convincing, because there remains the possibility that the large-angle anisotropies were generated more recently, as the microwave photons propagated towards us. A class of theories capable of doing this are those where structure is induced by topological defects such as cosmic strings [9] or textures [10]. Such theories are technically much more complicated, and consequently their predictions, to be set against a host of large scale structure observations, have not been established to nearly the same extent as more widely investigated inflation-based models such as the Cold Dark Matter model [11] which do rely on the existence of large-scale density perturbations. However, future observations and/or theoretical developments are certainly capable of excluding or supporting such models.

## II. THE RECENT UNIVERSE

We believe that the evolution of the universe as a whole is well understood from nucleosynthesis onwards. It can be described via an isotropic metric

$$ds^2 = -dt^2 + a^2(t)\left[\frac{dr^2}{1-kr^2} + r^2 d\theta^2 + r^2\sin^2\theta\, d\phi^2\right], \tag{1}$$

where $a(t)$ is the scale factor, $k$ is a constant measuring the spatial curvature and the speed of light is set to unity throughout. The main uncertainties are the present expansion rate, given by the Hubble parameter $H = \dot{a}/a$ with overdots being time derivatives, and the present density of matter (as a fraction of the critical density) $\Omega$ which governs if the universe is open, closed or spatially flat. If one extrapolates into the past the density rapidly approaches that giving a spatially flat universe, so our present knowledge that it is within an order of magnitude of flatness allows us to neglect the spatial curvature at the early times that this paper focuses on. The present value of the Hubble parameter shall, as usual, be parametrized as $H = 100h$ km s$^{-1}$Mpc$^{-1}$ = $h/3000$ Mpc$^{-1}$ with $h$ conservatively constrained to the range $h \in [0.3, 1]$.

Times are conventionally given in terms of redshift $z$ defined by $1 + z = a(t_0)/a(t)$, where $t_0$ is the present time and $a(t_0)$ can be normalized to unity. This means that present physical distances coincide with comoving distances, the latter being distances measured in coordinates dragged along with the expansion. Comoving distances shall be used throughout this paper.

The present universe is matter dominated with the density of radiation accurately given via the microwave background temperature of around 3K. Extrapolating backward, the universe became radiation dominated at a redshift $z_{\rm eq} \simeq 24000\Omega h^2$. The decoupling of the microwave background radiation occurs at $z_{\rm dec} \simeq 1000$, almost independently of $\Omega$ or $h$. Considerably before either of these times is the time of nucleosynthesis [4], which occurred at a temperature of around $10^{10}$K. The abundances of light elements are very sensitive to the expansion rate at that time, providing a very accurate measure of the Hubble parameter then and confirming the view that the evolution of the universe at least from that time onwards is well understood.

Making the excellent approximation that the transition from radiation domination to matter domination is instantaneous, we can easily obtain the comoving Hubble distance $H^{-1}/a$ at both decoupling and nucleosynthesis

$$\frac{H^{-1}_{\rm dec}}{a_{\rm dec}} = 95h^{-1}{\rm Mpc}\,; \tag{2}$$

$$\frac{H^{-1}_{\rm nuc}}{a_{\rm nuc}} = 10^{-4}\,{\rm Mpc}\,. \tag{3}$$

The Hubble length is the characteristic scale of an expanding universe, and is important because under normal circumstances such as matter or radiation domination it provides a good estimate of the distance light can travel. This communication distance, in comoving units, that light can travel between two times is

$$d_{\rm comm}(t_1, t_2) = \int_{t_1}^{t_2} \frac{dt}{a(t)}\,. \tag{4}$$

The concept of communication is closely tied to the ability to create a density perturbation in an expanding universe; to establish a density perturbation on a given scale is equivalent to sending a communication on the scale of the perturbation. Imposing causality therefore ensures that the communication distance limits the scale on which perturbations can be generated.

## III. THE LIMIT TO CAUSALITY

### A. The inflationary universe

The inflationary universe is defined by the condition that the scale factor is accelerating, $\ddot{a} > 0$. This is precisely equivalent to a *decreasing* comoving Hubble length, $d(H^{-1}/a)/dt < 0$. This demonstrates how inflation can generate density perturbations. Originally the comoving Hubble length is large, enabling perturbations to be generated causally. The inflationary epoch then greatly decreases the comoving Hubble length, to such an extent that even its subsequent growth after inflation ends is insufficient to make it as large as its pre-inflationary value. Communication does not play a role here once the perturbations are set up; they can just remain fixed in comoving coordinates and wait for the Hubble length to shrink past them.

The true power of inflation is not just that it permits super-Hubble-radius perturbations to be formed, but that it provides a specific and unambiguous mechanism via the stretching of quantum fluctuations [7].

At the same time, inflation has often been challenged through the lack of a convincing specific model. To a large extent the 'naturalness' issue has been superseded; the quality of large scale structure observations is now such that one anticipates that the question of whether or not inflation occurred can be addressed directly from observations, rather than from a philosophical standpoint.

We have just seen how inflation permits the generation of super-Hubble-radius perturbations through the decrease of the comoving Hubble length. This being the defining property of inflation, the question arises if this is the only possible means of doing so. Without inflation the comoving Hubble length increases monotonically, so matter must be moved to generate super-Hubble-radius perturbations and the communication distance becomes crucial.

### B. Communication in general relativity

Regardless of its nature, matter in an isotropic universe can be described by a fluid with an energy density $\rho$ and a pressure $p$, provided one is willing to accept an arbitrary time dependence for the pressure. Recalling that we can treat the universe as spatially flat, the equations of motion are

$$H^2 = \frac{8\pi}{3m_{\rm Pl}^2}\rho\,;\qquad(5)$$

$$\dot{\rho} = -3H\left(\rho+p\right)\,,\qquad(6)$$

where $m_{\rm Pl}$ is the Planck mass. These can be combined into the acceleration equation

$$\frac{\ddot{a}}{a} = -\frac{4\pi}{3m_{\rm Pl}^2}\left(\rho+3p\right)\,.\qquad(7)$$

By assumption we are forbidding inflation, so at all times we must have $p \geq -\rho/3$.

Large communication distances originate from small scale factors. The first result we need to establish is how rapidly $a$ can decrease into the past as a function of the density. Adopting the density as a time variable gives the elegant equation

$$\frac{d\ln a(\rho)}{d\ln\rho} = -\frac{1}{3}\left(\frac{\rho}{\rho+p}\right)\,,\qquad(8)$$

from which clearly the most rapid decline of $a(\rho)$ as $\rho$ increases corresponds to the lowest possible pressure $p = -\rho/3$. (This is also clear from simple thermodynamic grounds of work done against the expansion.) This corresponds to the 'coasting' solution $a(t) \propto t$, $\rho \propto a^{-2}$.

We now wish to find the maximum communication distance achievable as the universe evolves *without inflation* between two densities. The reason that densities are appropriate is because the limit to causal evolution is set by the Planck density $\rho_{\rm Pl}$, above which quantum gravity becomes important and the very notion of causality presumably breaks down. Using the density as the integration variable in the communication formula gives

$$d_{\rm comm}(t_1,t_2) = -\frac{m_{\rm Pl}}{\sqrt{24\pi}}\int_{\rho_1}^{\rho_2}\frac{1}{a(\rho)}\frac{1}{(\rho+p)}\frac{1}{\sqrt{\rho}}\,d\rho\,.\qquad(9)$$

This can be maximized by separately maximizing the terms in the integrand. Recalling that $a(\rho)$ is normalized at the present and the extrapolation back to nucleosynthesis is assumed, the value of $a(\rho_{\rm nuc})$ is fixed. The solution where $a(\rho)$ declines most rapidly with increasing $\rho$ therefore maximizes the first term. The second term in the integrand is maximized by the lowest pressure; as it happens this coincides with the condition that maximizes the first term. Hence the 'coasting' evolution[*] maximizes the communication distance between the Planck era and nucleosynthesis.

With the coasting evolution, the ratio of the maximum communication distance to the comoving Hubble length at nucleosynthesis is

$$\frac{d^{\rm max}_{\rm comm}(\rho_{\rm Pl},\rho_{\rm nuc})}{H^{-1}_{\rm nuc}/a_{\rm nuc}} = \frac{1}{2}\ln\frac{\rho_{\rm Pl}}{\rho_{\rm nuc}}\,.\qquad(10)$$

Since $\rho_{\rm Pl} \sim (10^{19}\,{\rm GeV})^4$ and $\rho_{\rm nuc} \sim (10^{-3}\,{\rm GeV})^4$, the logarithmic factor is around 200. By nucleosynthesis, the maximum comoving communication distance is therefore 0.01 Mpc, a small fraction of the Hubble radius at decoupling.

To complete the argument, we need to know about the communication distance between nucleosynthesis and decoupling. These two events being either side of matter-radiation equality, the distance is bounded above by assuming matter domination throughout that interval to obtain

$$\frac{d^{\rm max}_{\rm comm}(\rho_{\rm nuc},\rho_{\rm dec})}{H^{-1}_{\rm dec}/a_{\rm dec}} \leq 2\,.\qquad(11)$$

Putting these pieces together, the maximum possible communication distance that can be achieved between the Planck era and decoupling *without inflation* is less than $200h^{-1}$Mpc. This distance subtends an angle of less than $2°$ on the microwave sky.

### C. Communication in extended gravity theories

The same argument goes through in every known extended gravity theory which possesses a metric, including scalar-tensor theories and higher-order gravity theories (see Ref. [12] for extensive discussion), with only one difference. The reason is that the whole argument depends only on the behaviour of the scale factor. Whatever the true equations of motion might be, one can always *define* an 'effective' density $\tilde{\rho}$ by

$$H^2 = \frac{8\pi}{3m_{\rm Pl}^2}\tilde{\rho}, \qquad (12)$$

where $m_{\rm Pl}$ is now the present-day Planck mass. That is, the density is defined from the derivative of the scale factor. Similarly an effective pressure $\tilde{p}$ is defined by

$$\dot{\tilde{\rho}} = -3H\left(\tilde{\rho} + \tilde{p}\right), \qquad (13)$$

which amounts to defining the effective pressure from the second derivative of the scale factor. Whether or not these have anything to do with the actual pressure or density is irrelevant, and the influence of any further equations of motion is incorporated in the arbitrary time dependence of $\tilde{p}$.

---

[*]In general relativity coasting evolution can arise in a number of ways. A universe dominated either by curvature or by strings coasts, but it is hard to see how the curvature 'energy' or string energy could later be converted back into more useful forms of matter. More attractive is a scalar field with a suitably chosen exponential potential, whose decay may provide 'reheating' to a more conventional universe. Regardless, we shall see that even coasting evolution is not good enough.

Crucially, the derivations of Eqs. (7) and (9) remain the same in terms of these new effective quantities, and so the coasting solution remains the most effective non-inflationary evolution for maximising the communication distance. From nucleosynthesis onwards $\tilde{\rho}$ and $\tilde{p}$ must coincide with $\rho$ and $p$ respectively.

The only difference, alluded to above, is that in extended gravity theories the Planck mass may be varying, changing the definition of $\tilde{\rho}_{\rm Pl}$. The effective Planck mass $M_{\rm Pl}(t)$ governs the coupling of matter to gravity, and as usual it is defined by the coefficient of the Einstein-Hilbert term in the Lagrangian being $M_{\rm Pl}^2(t)/16\pi$. The archetypal example is a scalar-tensor theory [12], where $M_{\rm Pl}^2(t)$ is given by the value of the Brans-Dicke field. The quantum corrections generated by such a term include the curvature invariants $R^2$, $R_{\mu\nu}R^{\mu\nu}$ and $R_{\mu\nu\rho\sigma}R^{\mu\nu\rho\sigma}$, with dimensionless coefficients assumed to be of order unity. The condition that the quantum gravity corrections are important[†] is therefore that at least one of the curvature invariants exceeds $M_{\rm Pl}^4(t)$.

For isotropic universes, the condition that at least one of these invariants is of the order $M_{\rm Pl}^4(t)$ is that $\tilde{\rho}_{\rm Pl} \simeq m_{\rm Pl}^2 M_{\rm Pl}^2$, the appearance of the present-day value $m_{\rm Pl}$ being a quirk of the definition of $\tilde{\rho}$. Eq. (10) becomes

$$\frac{d_{\rm comm}^{\rm max}(\tilde{\rho}_{\rm Pl}, \tilde{\rho}_{\rm nuc})}{H_{\rm nuc}^{-1}/a_{\rm nuc}} = 100 + \ln \frac{M_{\rm Pl}(t_{\rm Pl})}{m_{\rm Pl}}. \qquad (14)$$

Recall that this ratio has to be greater than $10^6$ for the communication scale at nucleosynthesis to remain much greater than the Hubble radius up to decoupling. So even within this generalized gravity scenario, such perturbations can only be generated if the Planck mass has increased to at least $10^{500,000}$ GeV by the time the Planck scale is achieved! It seems likely that if models exist where it can reach so high, those models will have a Planck mass which diverges fast enough that the Planck scale is never reached in the universe's past. Such a scenario is the only loophole to the argument presented here.

## IV. CONCLUSION

The simplicity of the arguments presented here is because the question of perturbation generation is constrained directly by the kinematics of the Robertson-Walker space-time; although pressures and densities have been mentioned the entire argument could have been phrased without them. Because of this, the result is of considerable generality, applying in all metric theories of gravitation.

The closest cousin to the present paper is that of Hu, Turner and Weinberg [3], who discussed whether inflation was the unique way of solving the horizon and flatness problems. The flatness problem has not been discussed in this paper; note though that the condition for inflation is precisely the condition that the universe approaches spatial flatness rather than diverges from it, so inflation is clearly a necessary condition — Ref. [3] shows that entropy production is also required. The horizon problem, on the other hand, is essentially equivalent to the ability to generate density perturbations as they both rely on communication. This paper can therefore be regarded as a considerably simpler proof of the same result as in Ref. [3]. It is more general, since although parts of their argument apply in arbitrary metric theories of gravity, there are parts specific to general relativity and parts specific to scalar-tensor theories. Further, they exclude consideration of models close to the 'coasting' solution as a loophole to their analysis, whereas in the present paper these cases are also shown to be ineffective in allowing density perturbation generation. The rephrasing in terms of perturbation generation also seems more striking though that is a matter for personal prejudice.

Although at present theories based on topological defects provide examples of theories capable of generating large angle microwave background fluctuations *without* ever generating perturbations well above the Hubble radius, it is certainly very possible that in the near future only models relying on super-Hubble-radius perturbations will prove viable. Should this prove to be the case, the fact that there are such perturbations at all is in many ways as interesting as whether or not their form might follow a standard pattern such as the Harrison-Zel'dovich spectrum. Regardless of the underlying gravitational theory, one would be compelled to accept that the perturbations are due either to the acausal processes of quantum gravity, or that a period of cosmological in-flation must have occurred.

---

[†]The quantum corrections considered here are only those generated from the Einstein-Hilbert term, and not from any other gravitational term which may be present in the action.

## ACKNOWLEDGEMENTS

ARL is supported by the Royal Society, and thanks John Barrow, Ed Copeland, Jim Lidsey, David Lyth and especially David Wands for helpful discussions.

---

[1] E. W. Kolb and M. S. Turner, *The Early Universe* (Addison-Wesley, Redwood City, CA, 1990); A. D. Linde, *Particle Physics and Inflationary Cosmology* (Harwood Academic, Chur, Switzerland, 1990); A. R. Liddle and D. H. Lyth, Phys. Rep **231**, 1 (1993).
[2] G. F. Smoot et al., Astrophys. J. Lett. **396**, L1 (1992); C. L. Bennett et al., "Cosmic Temperature Fluctuations from Two Years of COBE DMR Data", to appear, Astrophys. J. (1994).
[3] Y. Hu, M. S. Turner and E. J. Weinberg, Phys. Rev. D**49**, 3830 (1994).
[4] T. Walker, G. Steigman, D. N. Schramm, K. A. Olive and H.-S. Kang, Astrophys. J. **376**, 51 (1991).
[5] D. Scott, J. Silk and M. White, "From Microwave Anisotropies to Cosmology", Berkeley preprint (1994).
[6] A. H. Guth, Phys. Rev. D**23**, 347 (1981).
[7] A. H. Guth and S.-Y. Pi, Phys. Rev. Lett. **49**, 1110 (1982); S. W. Hawking, Phys. Lett. **B115**, 339 (1982); A. A. Starobinsky, Phys. Lett. **B117**, 175 (1982).
[8] See eg A. D. Linde in Ref. [1].
[9] A. Vilenkin and E. P. S. Shellard, *Cosmic Strings and Other Topological Defects* (Cambridge University Press, Cambridge, 1994).
[10] U.-L. Pen, D. N. Spergel and N. Turok, Phys. Rev. D**49**, 692 (1994).
[11] M. Davis, G. Efstathiou, C. S. Frenk and S. D. M. White, Nature 356, 489 (1992) and refs therein.
[12] C. M. Will, *Theory and Experiment in Gravitational Physics* (Cambridge University Press, Cambridge, 1993).